\documentstyle[twocolumn,prl,aps,epsfig,floats]{revtex}
\begin{document}




{\bf Few comments on preprint cond-mat/0007430 by Aleiner et al.:}
It has been suggested in Ref.~\cite{ZarZaw} that conduction electron 
generated virtual hopping to  the excited states of a two-level system (TLS)
could increase the amplitude of assisted tunneling and thus result in a 
measurable Kondo temperature. In a recent preprint, Aleiner {\em et  al.} 
revisited the role of excited states\cite{Aleiner} and 
pointed out that in the calculations of Ref.~\cite{ZarZaw}
only the first few excited states have been taken into account and 
inclusion of {\em all} excited states induces a large decrease 
of $T_K$ instead of  an increase. 

While we think that the work of Aleiner {\em et al.} addresses an important 
issue, there are several points that we have to comment on: 

{\em (a)} The reduction of $T_K$ heavily relies on the assumption 
of {\em electron-hole} symmetry. In fact, using the model of 
Ref.~\cite{ZarZaw} and assuming a contact TLS-electron 
interaction of strength $U$, the assisted tunneling 
term generated by the reduction of the cutoff is proportional to:
\begin{equation}
\delta V^{\alpha\beta}_{ll'}  =  
\sum _{\epsilon_j < D} \sum_{l''} \varrho(D) V^{\alpha j}_{ll''}
V^{j\beta}_{l''l'} 
 -  \varrho(-D) V^{j\beta}_{ll''} V^{\alpha j}_{l''l'},
\label{eq:scaling}
\end{equation}
where $D$ denotes the bandwidth cutoff, $\epsilon_j$ is the energy of 
the j'th  excited state of the TLS, and the matrix element between 
the TLS wave functions $\varphi_i$, $\varphi_j$ in the "tower" model 
\cite{Aleiner} and conduction electron 
states $j_l(z)$ $j_{l'}(z)$ is given by:
\begin{equation}
 V^{i j}_{ll'} \sim U \int dz \;\varphi_i(z)\varphi_j(z) j_l(z) j_{l'}(z)\;.
\end{equation}
For large enough $D$'s the identity $\sum \varphi_j(z)\varphi_j(z') \approx
\delta(z-z')$ can be used and assuming  $\varrho(D) = \varrho(-D)$ the two 
terms in Eq.~(\ref{eq:scaling}) cancel. However, for electronic bands usually 
$\varrho(D)\ne
\varrho(-D)$ and the cancellation does not occur. 
In the Figure we show $T_K$  obtained from the complete 
solution of the leading logarithmic scaling equations. We assumed 
a density of states of the form 
$$
\varrho(\epsilon) = \varrho(0)\bigl(1+\alpha \;{\epsilon/ D_0}\bigr)\;,
$$
where $\alpha$ is usually of the order of 1. As one can see, a small
electron-hole asymmetry can increase $T_K$ by many orders of  magnitude, 
bringing it into the physically measurable range.  
Note that while for $\alpha = 0$ the inclusion of all excited states 
reduces $T_K$ by three orders of magnitude in agreement with 
Aleiner {\em et al.}, for $\alpha = 0.3-0.4$ there is very little reduction 
compared to the truncated model. [The $\alpha=0.4$ case is special: Here
five  heavy  particle states are still active at $T_K$.] 

{\em (b)} It is obvious from the figure that it is extremely 
difficult to give a 'first principles' estimation for $T_K$, which 
is very sensitive to the exact model parameters such as the 
distance between the two wells, the mass of a heavy particle, 
or the  value and structure of the heavy particle-conduction 
electron interaction. Within the Thomas-Fermi interaction 
approximation, e.g., $T_K$ changes many orders of magnitudes if one 
changes the effective charge of the heavy particle by a factor of 2
(Hall effect measurements indicate, e.g., that the valence of crystalline 
$Cu$  is about $1.5$. This value may considerably differ from the valence 
of the tunneling atoms that are most probably sitting in a 
more disordered region.)

In reality, one does not even know what  tunneling centers 
look like. The simplistic model of Ref.~\cite{ZarZaw} assumes a very 
specific TLS structure, which may be very far from that 
of real tunneling systems, possibly due to dislocation  
jogs, e.g.\cite{Ralph}.  

{\em (c)} Our calculations clearly demonstrate that $T_K$ {\em can 
be in the measurable range}. It is a more delicate  question 
whether the splitting of the TLS can be small enough to 
allow for the development of a two-channel Kondo behavior. 
In our model the bare splitting is $\Delta_0 \sim 0.4K$, which is 
clearly smaller  than  $T_K$ for $\alpha > 0.2$. 
However, the presence of electron-hole symmetry  breaking
{\em increases} $\Delta_0$, which is, on the other hand  
{\em considerably reduced} under scaling. To decide 
which of these processes wins and whether the renormalized splitting 
can be less than $T_K$,  a much more detailed 
study of  the next leading logarithmic equations is needed. 

Another possibility could  be to have  tunneling centers, where the 
degeneracy of the lowest-lying states is guaranteed by some local 
symmetry (rotational or possibly time reversal). 

\noindent
A. Zawadowski${}^1$ and G. Zar\'and,$^{1,2}$ \\
{\small
${}^1$Institute of Physics, Technical University of Budapest,
$^2$Lyman Physics Laboratory, Harvard University
}

\begin{figure}[b]
\begin{center}
\psfig{figure=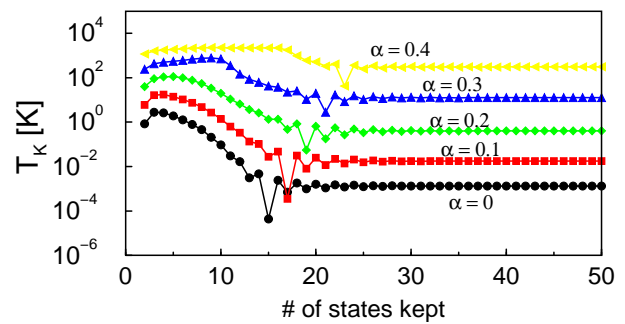,width=8cm}
\end{center}
\vspace*{0.05cm}
\caption{
$T_K$ as a function of the excited states kept in the scaling procedure.
The calculations were done for assuming a heavy particle of mass 
$M\sim 50 \times m_{\rm proton}$, a barrier of  height $\sim 300 K$
and of width  $0.5\AA$, and the width of the potential wells
was taken to be  $0.1\AA$. 
}
\end{figure}

\end{document}